%% file: zpt.tex
\documentclass[aps,prl,showpacs,twocolumn,groupedaddress]{revtex4}  % for submission
\usepackage{graphicx}  % needed for figures
\usepackage{dcolumn}   % needed for some tables
\usepackage{bm}        % for math
\usepackage{amssymb}   % for math

\def\met{\mbox{${\hbox{$E$\kern-0.6em\lower-.1ex\hbox{/}}}_T$}} %missing ET

\newcommand{\NCCCC}{23,959}
\newcommand{\NCCEC}{30,344}
\newcommand{\NECEC}{9,598}
\newcommand{\bckcccc}{1.30} % percent
\newcommand{\bckccccerr}{0.14} 
\newcommand{\bckccec}{8.55} % percent
\newcommand{\bckccecerr}{0.26} % percent
\newcommand{\bckecec}{4.71} % percent
\newcommand{\bckececerr}{0.30} % percent
\newcommand{\chicssndf}{11} 
\newcommand{\chicss}{11.1} 
\newcommand{\chimod}{31.9} 
 
\newcommand{\acceptptzero}{0.27} 
\newcommand{\acceptmin}{0.19} 
\newcommand{\ptacceptmin}{40}

\newcommand{\zgamma}{$Z$}
\begin{document}

\hspace{5.2in} \mbox{Fermilab-Pub-07/642-E}
\title{
 Measurement of the shape of the boson transverse momentum 
distribution in  $p\bar{p} \rightarrow Z /\gamma^{*} \rightarrow e^+e^- +X $ events
produced at $\sqrt{s}=1.96$ TeV
}
\input list_of_authors_r2.tex  % input Dzero author list
\date{December 5, 2007}

\begin{abstract}
We present a measurement of the shape of the $Z/\gamma^{*}$ boson transverse
momentum ($q_T$) distribution in $p\bar{p}\rightarrow Z/\gamma^{*}\rightarrow e^+e^-+X$
events at a center-of-mass energy of 1.96 TeV using 0.98 fb$^{-1}$ of data collected
with the D0 detector at the Fermilab Tevatron collider.
The data are found to be consistent with the resummation prediction 
at low $q_T$, but above the perturbative QCD calculation in
the region of $q_T>30$ GeV/$c$. Using events with $q_T<30$ GeV/$c$, we 
extract the value of $g_2$, one of the non-perturbative parameters for the 
resummation calculation. Data at large boson rapidity $y$ are compared with the 
prediction of resummation and with alternative models that employ a resummed 
form factor with modifications in the small Bjorken~$x$ region of the proton wave function.
\end{abstract}

\pacs{13.60.Hb, 13.38.Dg, 13.85.Qk, 13.38.Bx}
\maketitle

\input{intro.tex}
\input{detector.tex}

\input{analysis.tex}

\input{result.tex}
\input{conclusion.tex}

\input acknowledgement_paragraph_r2.tex   % input acknowledgement

\end{document}

%% file: list_of_authors_r2.tex
% LIST_OF_AUTHORS_R2.TEX               10/09/07(b)          
%
\author{V.M.~Abazov$^{36}$}
\author{B.~Abbott$^{76}$}
\author{M.~Abolins$^{66}$}
\author{B.S.~Acharya$^{29}$}
\author{M.~Adams$^{52}$}
\author{T.~Adams$^{50}$}
\author{E.~Aguilo$^{6}$}
\author{S.H.~Ahn$^{31}$}
\author{M.~Ahsan$^{60}$}
\author{G.D.~Alexeev$^{36}$}
% ...

\affiliation{\vspace{0.1 in}(The D\O\ Collaboration)\vspace{0.1 in}}
% ...
\affiliation{$^{6}$University of Alberta, Edmonton, Alberta, Canada,
                Simon Fraser University, Burnaby, British Columbia, Canada,
                York University, Toronto, Ontario, Canada, and
                McGill University, Montreal, Quebec, Canada}
% ...
\affiliation{$^{29}$Tata Institute of Fundamental Research, Mumbai, India}
% ...
\affiliation{$^{31}$Korea Detector Laboratory, Korea University, Seoul, Korea}
% ...
\affiliation{$^{36}$Joint Institute for Nuclear Research, Dubna, Russia}
% ...
\affiliation{$^{50}$Florida State University, Tallahassee, Florida 32306, USA}
% ...
\affiliation{$^{52}$University of Illinois at Chicago,
                Chicago, Illinois 60607, USA}
% ...
\affiliation{$^{60}$Kansas State University, Manhattan, Kansas 66506, USA}
% ...
\affiliation{$^{66}$Michigan State University,
                East Lansing, Michigan 48824, USA}
% ...
\affiliation{$^{76}$University of Oklahoma, Norman, Oklahoma 73019, USA}
% ...

%% file: intro.tex
A complete understanding of weak vector boson production is essential for maximizing
the sensitivity to new physics at hadron colliders. %through precision measurements of the $W$ boson mass,
%detailed studies of top quark production, and searches for
%production of the Higgs boson and other phenomena beyond the standard model.
Studies of the $Z/\gamma^{*}$ boson production play a particularly valuable role in that
its kinematics can be precisely determined through measurement of its leptonic decays. 
Throughout this Letter, we use the notation ``$Z$ boson'' to mean ``$Z/\gamma^{*}$ boson'', 
unless specified otherwise.

\zgamma ~boson production also serves as an ideal testing ground
for predictions of quantum chromodynamics (QCD), since the boson's
transverse momentum, $q_T$, can be measured over
a wide range of values and can be correlated with its
rapidity. At large $q_T$ (approximately greater than 30 GeV/$c$), 
the radiation of a single parton with large transverse momentum dominates the cross
section, and fixed-order perturbative QCD (pQCD) calculations \cite{pQCD, melnikov}, 
%currently available at next-to-next-to leading order (NNLO) \cite{melnikov},
should yield reliable predictions. At lower $q_T$,
multiple soft gluon emission can not be neglected,
%and pQCD calculation no longer gives accurate results.
and the fixed-order perturbation calculation no longer gives accurate results.
A soft-gluon resummation technique developed by Collins, Soper, and Sterman
(CSS) \cite{CSS} gives reliable predictions in the low-$q_T$ region.
A prescription has been proposed \cite{Arnold} for
matching the low- and high-$q_{T}$ regions in order to provide a continuous
prediction for all values of $q_{T}$. The CSS resummation formalism allows the inclusion of
contributions from large logarithms of the form $\ln^n(q^{2}_T/Q^{2})$ to all orders of perturbation theory in an effective
resummed form factor, where $Q^2$ represents the invariant mass corresponding to the four-momentum transfer.
The CSS resummation can be done either in impact parameter ($b$) space or in transverse momentum ($q_T$) space. 
In the case of $b$-space resummation, this form factor can be parameterized with the following 
non-perturbative function first introduced by Brock, Landry, Nadolsky and Yuan (BLNY) \cite{Yuan2}:
\begin{equation}
S_{NP}(b, Q^{2}) = \left[g_{1} + g_{2} \ln \left( \frac{Q}{2Q_{0}} \right) +g_1g_3\ln(100 x_i x_j)\right] b^{2},
\end{equation}
where $x_i$ and $x_j$ are the fractions of the incident hadron momenta carried by 
the colliding partons, $Q_{0}$ is a scale typical of the onset of non-perturbative effects, 
and $g_1$, $g_2$ and $g_3$ are phenomenological non-perturbative parameters that must be obtained from 
fits to the data. The $Z$ boson $q_T$ distribution at the Fermilab Tevatron is by far most sensitive to the 
value of $g_2$ and quite insensitive to the value of $g_3$. 
Thus a measurement of the \zgamma~boson $q_T$ spectrum can be used to test this
formalism and to determine the value of $g_2$.

Recent studies of data from deep inelastic scattering
(DIS) experiments \cite{DIS1, DIS2} indicate that the resummed form factor in the above equation may need
to be modified for processes involving a small-$x$ parton in
the initial state. Ref.~\cite{small x} indicates how such a modification
would influence the $q_T$ distributions of vector and Higgs bosons
produced in hadronic collisions. A wider $q_T$ distribution is predicted
%produced in hadronic collisions. A wider transverse momentum distribution is predicted
for \zgamma~bosons with large rapidity (called ``small-$x$ broadening'').
\zgamma~bosons produced at the Tevatron in the rapidity range $2<|y|<3$ 
probe processes involving a parton with $0.002<x<0.006$, 
and can be used to test the modified form factor at small $x$.
%probe processes involving a parton with Bjorken $x$ between 0.002 and 0.006, 
%and can be used as a test of the modified resummed form factor at small $x$.

\zgamma~boson $q_T$ distributions have been published previously
by the CDF \cite{cdfruni} and D0 \cite{d0runi} collaborations using
about 100 pb$^{-1}$ of data at $\sqrt{s}=1.8$ TeV. In this Letter, we report a new measurement
%about 100 pb$^{-1}$ of data at a center-of-mass energy of 1.8 TeV. In this Letter, we report a new measurement
with larger statistics and improved precision. 
This measurement is also the first to present a $q_{T}$ distribution for large-rapidity $Z$ bosons.

%% file: detector.tex
The data sample used in this measurement was collected using a set of 
inclusive single-electron triggers with the D0 detector \cite{d0det}
at the Fermilab Tevatron collider,
and the integrated luminosity is $980 \pm 60$ pb$^{-1}$ \cite{d0lumi}.
%at the Fermilab Tevatron collider at a center-of-mass energy of 1.96 TeV;
%the integrated luminosity of the data sample is $980 \pm 60$ pb$^{-1}$ \cite{d0lumi}.
%The D0 detector includes a central tracking system, composed of a
%silicon microstrip tracker and a central fiber tracker, 
%both located within a 2 T superconducting solenoidal magnet and
%optimized for tracking and vertexing capabilities at pseudorapidities
%of $|\eta|<3$ and $|\eta|<2.5$ respectively ($\eta=-\ln[\tan(\theta/2)]$, where
%$\theta$ is the polar angle with respect to the proton direction). 
%Three liquid argon and uranium calorimeters provide coverage out to $|\eta| \approx 4.2$:
%a central section with coverage of $|\eta|<1.1$ and two
%endcap calorimeters with an approximate coverage of $1.5<|\eta|<4.2$ for jets
%and $1.5<|\eta|<3.2$ for electrons. A muon system surrounds the 
%calorimetry and consists of three layers of scintillators
%and drift tubes and 1.8 T iron toroids with coverage of $|\eta|<2$. 

Our selection criteria for \zgamma~bosons require two
%Our selection criteria for candidate \zgamma~bosons require two
isolated electromagnetic clusters that have a shower shape consistent 
with that of an electron.
%with that of an electron and are away from the module boundaries of the calorimeters.
Electron candidates are required to have transverse momentum greater than
$25 ~\mbox{GeV}/c$. The electron pairs must have a reconstructed invariant mass 
$70<M(ee)<110 ~\mbox{GeV}/c^2$. If an event has both its candidate electrons in the
central calorimeter (CC events), each electron must be spatially matched to 
a reconstructed track. Because the tracking efficiency decreases with rapidity in the endcap region,
events with one or two endcap calorimeter electron candidates
(CE and EE events, respectively) are required to have at least one electron with a matching track.
After these requirements, \NCCCC ~CC, \NCCEC ~CE, and \NECEC ~EE
events are selected; 5412 of these have a reconstructed $Z$ boson with $|y|>2$.

%% file: analysis.tex
Electron identification efficiencies are measured using a combination
of data and a {\sc geant}-based \cite{geant} simulation of the D0 detector.
The electron identification efficiencies
are measured from \zgamma~data. 
%are measured from \zgamma~events 
%and are parameterized in terms of the electron
%transverse energy and, for some variables,
%the vertex position along the beam axis or electron incident angle.
The dependence of the overall selection efficiency on the \zgamma~boson $q_T$ is parameterized
from the {\sc geant} simulation. A measurement of this shape
from the data agrees well with the simulation within statistical uncertainties.

The dominant backgrounds are from photon plus jet events and
di-jet events, with photons and jets misidentified as electrons. The kinematic
properties of these events are obtained from events that satisfy most of the \zgamma~selection
criteria, but fail the electron shower shape requirement.
The normalization of the background is obtained by fitting 
to a sum of a signal shape obtained from a parameterized simulation 
of the detector response and the invariant mass distribution from the background sample
to the invariant mass distribution of the data sample. The background fractions are
(\bckcccc$\pm$\bckccccerr)\%,
(\bckccec$\pm$\bckccecerr)\%, and
(\bckecec$\pm$\bckececerr)\%~
for CC, CE, and EE events respectively. Other backgrounds are negligible.

The data are corrected for acceptances within a range of generated \zgamma~masses of 40 to 200 GeV/$c^{2}$,
and for selection efficiencies using a parameterized simulation. We
use {\sc ResBos} \cite{Yuan} as the event generator which incorporates the resummation calculation
in $b$-space using the BLNY parameterization for low $q_T$ and a NLO
pQCD calculation for high $q_T$. We use {\sc photos} \cite{photos} to simulate the effects of
final state photon radiation.
The overall acceptance times efficiency falls slowly from a value of
$\acceptptzero$ at low $q_T$ to a minimum of $\acceptmin$
at $q_T=$ \ptacceptmin ~GeV$/c$ and slowly increases for larger $q_{T}$.

The measured spectrum is further corrected for detector resolution effects using
the {\sc run} (Regularized Unfolding) program \cite{RUN} to obtain the true differential cross
section. Its performance was verified by comparing the true and unfolded
spectrum generated using pseudo-experiments. 
The measured $Z$ $q_T$ resolution is about 2 GeV/$c$; the bin width we choose 
is 2.5 GeV/$c$ for $q_T<30$ GeV/$c$. The typical correlation 
between adjacent bins is around 30\%. Due to limited statistics, the chosen bin width
is 10 GeV/$c$ for $30<q_T<100$ GeV/$c$ and 40 GeV/$c$ for $100<q_T<260$ GeV/$c$. 

Systematic uncertainties on the unfolded $q_{T}$ spectrum
arise from  uncertainties on the electron energy calibration,
the electron energy resolution, the dependence of the overall selection efficiency
on $q_T$, and the effect of parton distribution functions (PDFs) on the acceptance.
The uncertainties on the unfolded spectrum are estimated
from the resulting change when the smearing parameters are varied within their uncertainties. 
CTEQ 6.1M is used as the default PDF. Uncertainties due to the PDFs 
are estimated using the procedure described in Ref.~\cite{cteq}. The uncertainty due 
to the choice of unfolding parameters in the {\sc run} program
is also estimated and included in the final systematic uncertainty.

%% file: result.tex
\begin{figure*}[tbp]
\includegraphics[width=0.497\linewidth]{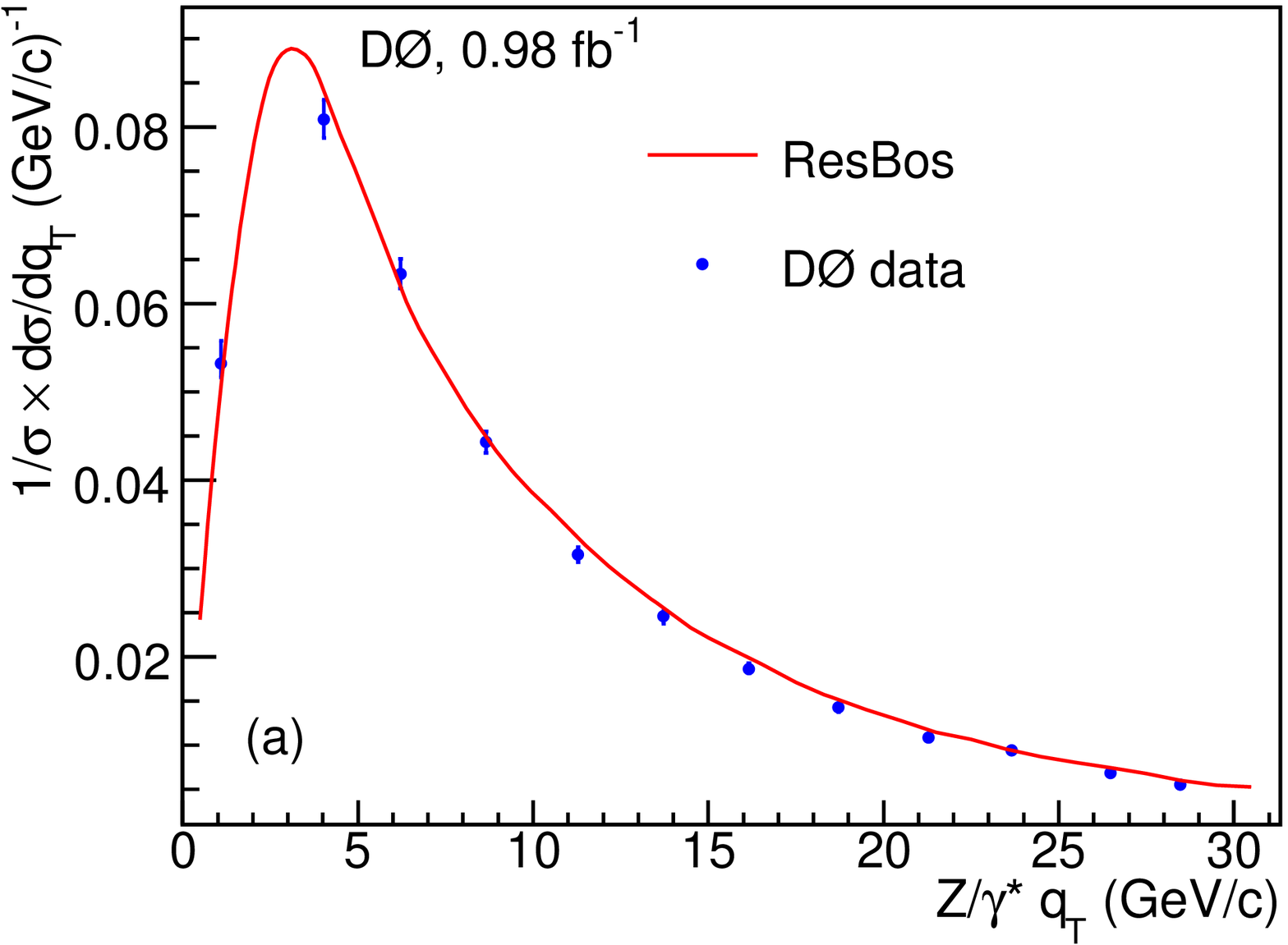}
\includegraphics[width=0.497\linewidth]{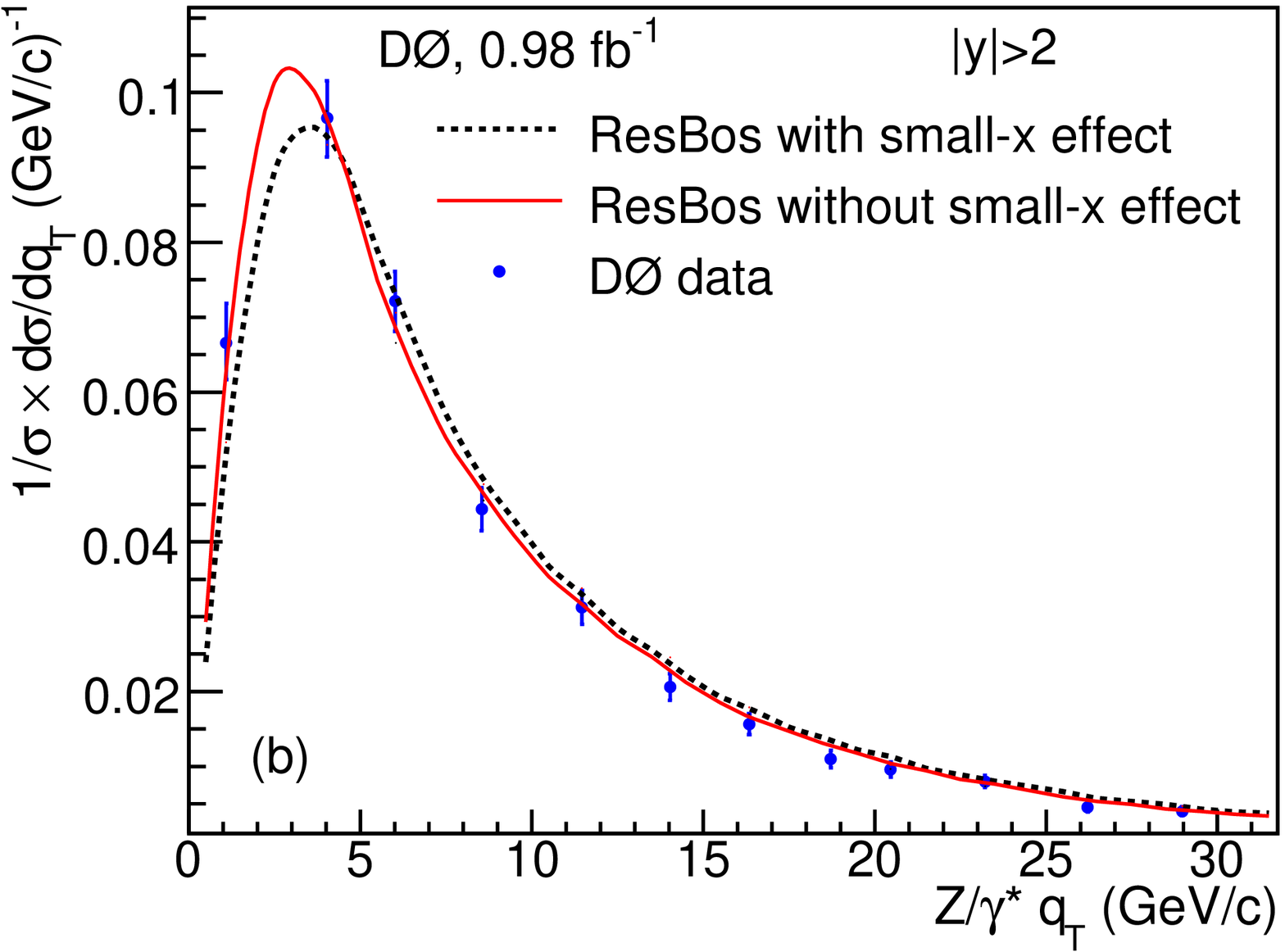}
\caption{\label{fig:finalresult}
The normalized differential cross section as a function of $q_{T}$
for (a) the inclusive sample and (b) the sample with \zgamma~boson $|y|>2$ with 
$q_T<30$ $\mbox{GeV}/c$. The points are the data, the solid curve 
is the {\sc ResBos} prediction, and the dashed line in (b) is the prediction
from the form factor modified after studies of small-$x$ DIS data. 
}
\end{figure*}

The final results in the $q_{T}<30$ GeV/$c$ range,
%with statistical and systematic uncertainties added in quadrature,
are shown in Fig.~\ref{fig:finalresult} for the inclusive
sample and for the sample with $|y|>2$. Each data point is 
plotted at the average value of the expected distribution over the bin \cite{Wyatt}.
For the theoretical calculation, we use {\sc ResBos} with published values of the 
non-perturbative parameters \cite{Yuan2}. Good agreement between data and the prediction 
%non-perturbative parameters \cite{Yuan2}. Good agreement between data and the {\sc ResBos} prediction 
is observed for all rapidity ranges, which indicates that 
the BLNY parameterization works well for the low $q_T$ region.

\zgamma~boson events produced at large rapidities ($|y|>2$) are also used to 
test the small-$x$ prediction. We compare data with the theoretical predictions 
with and without the form factor as modified from studies of small-$x$ DIS data \cite{small x}. 
All curves are normalized to 1 for $q_T<30$ GeV/$c$.
The default values for the parameters $g_{1}$, $g_{2}$, and $g_{3}$ \cite{Yuan2} obtained 
from large-$x$ data are used. The $\chi^2/$d.o.f.~between the data and the {\sc ResBos} 
calculation using the default parameters is 0.8/1 for $q_T<5$ GeV/$c$ and \chicss$/$\chicssndf~for $q_T<30$ GeV/$c$, 
while that for the modified calculation is 5.7/1 for $q_T<5$ GeV/$c$ and \chimod$/$\chicssndf~for $q_T<30$ GeV/$c$. 
It remains to be seen if retuning of the non-perturbative parameters could improve the agreement 
for the modified calculations.

\begin{figure*}[tbp]
\includegraphics[width=0.497\linewidth]{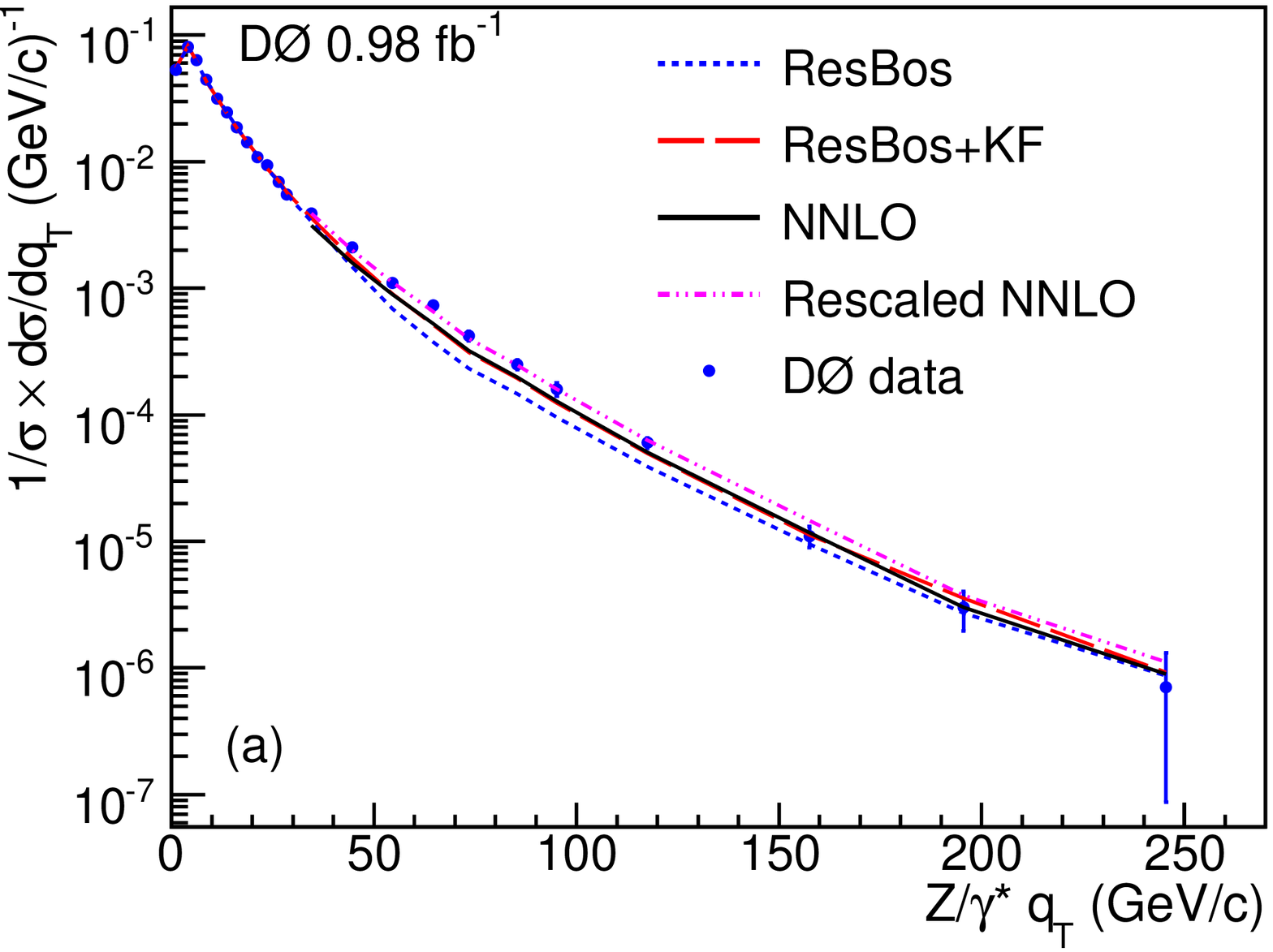}
\includegraphics[width=0.497\linewidth]{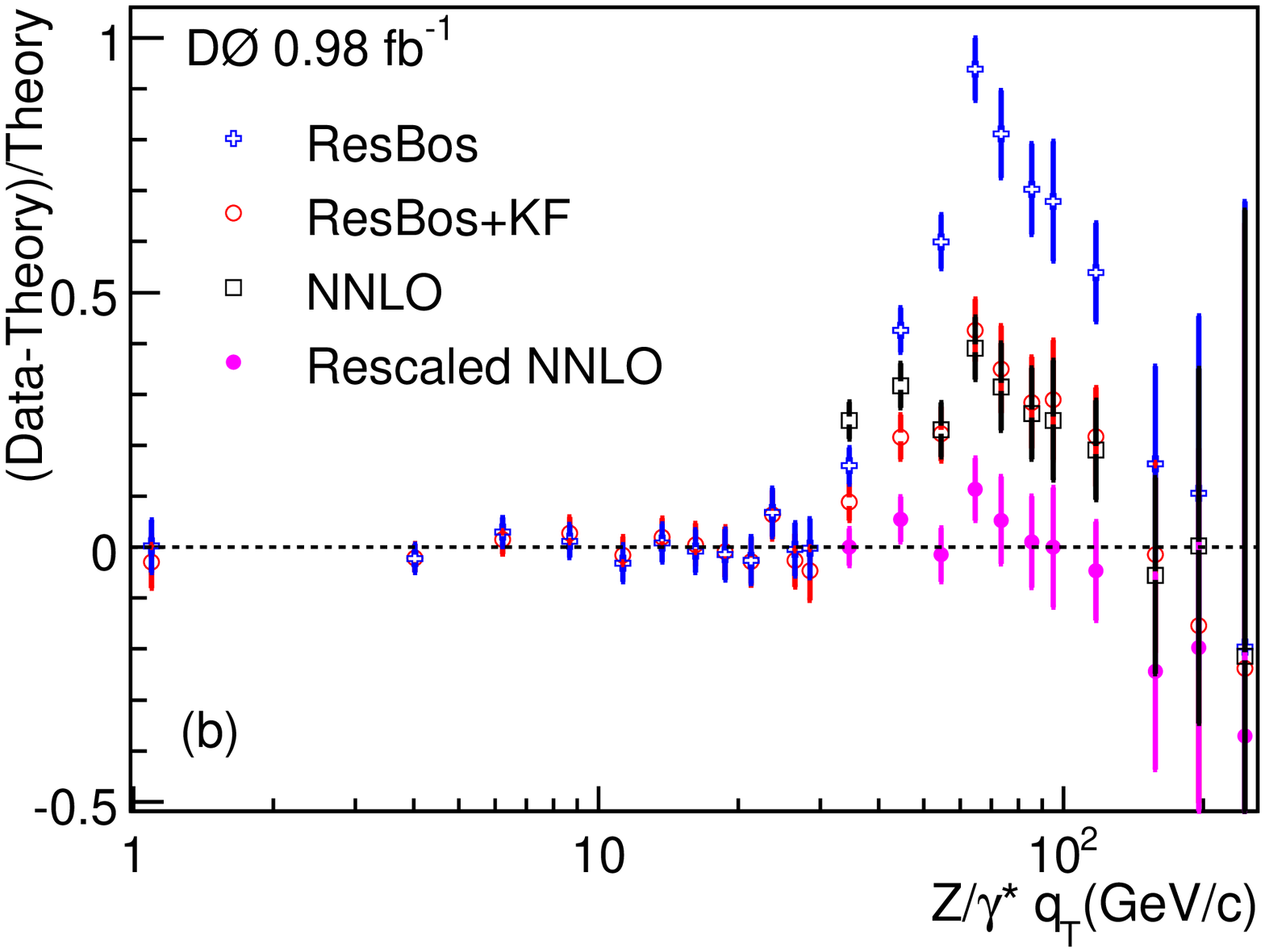}
\caption{\label{fig:finalresulty2}
The normalized differential cross section as a function of $q_{T}$ compared to four theoretical
calculations for (a) the entire range measured and (b) the fractional differences between
data and the theoretical predictions. The four theoretical calculations are {\sc ResBos}
with its default parameters, {\sc ResBos} with a NLO to NNLO K-factor by Arnold 
and Reno, the NNLO calculation by Melnikov and Petriello, and the NNLO calculation but 
rescaled to data at $q_T=30$ GeV/$c$.
}
\end{figure*}

Figure \ref{fig:finalresulty2} shows the measured differential cross section in
the range $q_{T}<260 ~\mbox{GeV}/c$ compared to (1) the {\sc ResBos} calculation with
its default parameters \cite{Yuan2}, (2) {\sc ResBos} with a NLO to NNLO K-factor
by Arnold and Reno \cite{ArnoldReno} incorporated into {\sc ResBos} by its authors, 
(3) a pQCD calculation at NNLO \cite{melnikov} using the MRST 2001 NNLO PDF set \cite{MRST}
divided by the NNLO calculation of the total cross section \cite{hamberg}, 
and (4) the NNLO calculation but rescaled to the data at $q_{T}=30 ~\mbox{GeV}/c$.
The agreement between data and {\sc ResBos}, with or without the K-factor, 
is good for $q_T<30 ~\mbox{GeV}/c$. At higher
%is good for values of $q_T$ less than about $30 ~\mbox{GeV}/c$. At higher
values of $q_T$, the data are not in agreement with the {\sc ResBos} calculation.
%The {\sc ResBos} calculation uses the result from the resummation calculation
%at low values of $q_T$, the purely perturbative result at high values of $q_T$, and a
%matching prescription at intermediate values.
The data agree better with the NNLO calculation and {\sc ResBos} prediction with the 
Arnold-Reno K-factor, but agrees best when the NNLO results are
rescaled by a factor of 1.25 so that they match the data at
$q_{T}=30~\mbox{GeV}/c$. This indicates that the shape from these calculations 
agrees with the data, and that the source of the discrepancy is in the normalization.
Table \ref{tab:theResult} summarizes the measured values for each $q_{T}$ bin together 
with statistical and systematic uncertainties.

The CSS model parameter most sensitive to the shape at low $q_T$ ($q_T<30$ GeV/$c$) is $g_2$.
In a fit, we fix other phenomenological parameters to the values obtained in Ref.~\cite{Yuan2}
and only vary $g_{2}$. A minimum $\chi^2$/d.o.f.~of $9/11$ between 
the model and the inclusive data for $q_{T}<30 ~\mbox{GeV}/c$ is found when $g_2=0.77 \pm 0.06$ ~(GeV/$c$)$^{2}$.

\begin{table}[!htb]
\begin{tabular}{c|c} \hline \hline
 $\langle q_T \rangle$ (GeV/$c$) & ~~~~~~~~ ${1/\sigma} \times d \sigma / d q_T$  {(GeV/$c$)$^{-1}$} ~~~~~~~ \\ \hline
  1.1  & $(5.32 \pm 0.13 \pm 0.24) \times 10^{-2}$ \\ 
  4.0  & $(8.08 \pm 0.12 \pm 0.19) \times 10^{-2}$ \\ 
  6.2  & $(6.33 \pm 0.11 \pm 0.14) \times 10^{-2}$ \\ 
  8.7  & $(4.43 \pm 0.09 \pm 0.11) \times 10^{-2}$ \\ 
  11.3 & $(3.15 \pm 0.08 \pm 0.08) \times 10^{-2}$ \\ 
  13.7 & $(2.46 \pm 0.07 \pm 0.06) \times 10^{-2}$ \\ 
  16.2 & $(1.86 \pm 0.06 \pm 0.05) \times 10^{-2}$ \\ 
  18.7 & $(1.42 \pm 0.05 \pm 0.05) \times 10^{-2}$ \\ 
  21.3 & $(1.09 \pm 0.04 \pm 0.03) \times 10^{-2}$ \\ 
  23.7 & $(9.40 \pm 0.40 \pm 0.20) \times 10^{-3}$ \\ 
  26.4 & $(6.90 \pm 0.30 \pm 0.20) \times 10^{-3}$ \\ 
  28.5 & $(5.50 \pm 0.30 \pm 0.10) \times 10^{-3}$ \\ 
  34.6 & $(3.90 \pm 0.10 \pm 0.10) \times 10^{-3}$ \\ 
  44.6 & $(2.10 \pm 0.07 \pm 0.06) \times 10^{-3}$ \\ 
  54.6 & $(1.10 \pm 0.05 \pm 0.03) \times 10^{-3}$ \\ 
  64.6 & $(7.30 \pm 0.40 \pm 0.20) \times 10^{-4}$ \\ 
  73.4 & $(4.20 \pm 0.30 \pm 0.20) \times 10^{-4}$ \\ 
  85.4 & $(2.50 \pm 0.20 \pm 0.10) \times 10^{-4}$ \\ 
  95.1 & $(1.60 \pm 0.17 \pm 0.08) \times 10^{-4}$ \\ 
  117.5 & $(6.00 \pm 0.50 \pm 0.30) \times 10^{-5}$ \\ 
  157.5 & $(1.10 \pm 0.20 \pm 0.07) \times 10^{-5}$ \\ 
  195.5 & $(3.00 \pm 1.00 \pm 0.30) \times 10^{-6}$ \\ 
  245.5 & $(7.10 \pm 6.10 \pm 0.60) \times 10^{-7}$ \\ \hline \hline 
\end{tabular}
\caption{
The normalized differential cross section for \zgamma~events produced in bins 
of $q_T$. 
The first uncertainty is statistical and the second is systematic. 
}
\label{tab:theResult}
\end{table}

%% file: conclusion.tex
In conclusion, we have measured the normalized differential spectrum,
${1 \over \sigma} {d \sigma \over d q_T}$, for \zgamma~boson events produced
in $p \bar{p}$ collisions at $\sqrt{s}=1.96$ TeV
%in $p \bar{p}$ collisions at a center-of-mass energy of 1.96 TeV
with boson mass $40<M<200$ GeV/$c^2$ and $q_T<260$ GeV/$c$. 
%with boson mass $40<M<200$ GeV/$c^2$ and transverse momentum $q_T<260$ GeV/$c$. 
This represents the highest center-of-mass energy measurement
of this quantity over the largest phase space available to date.
The overall uncertainty of this measurement
has been reduced compared with the previous measurements.
We find that for $q_T<30 ~\mbox{GeV}/c$, the CSS resummation model used in {\sc ResBos}
describes the data very well at all rapidities. Our data with $|y|>2$ disfavor a variant of this
model that incorporates an additional small-$x$ form factor when $g_{1}$, $g_{2}$, 
and $g_{3}$ from large-$x$ data is used.
Using the BLNY parameterization for events with $q_T<30$ GeV$/c$,
we obtain $g_{2}=0.77 \pm 0.06$ (GeV/$c$)$^{2}$, which is comparable with 
the current world average value \cite{Yuan2}.
We observe a disagreement between our data and NNLO calculations 
%We observe a disagreement between our data and NNLO calculations, 
%both those incorporated into {\sc ResBos} and a stand-alone calculation, 
in the region $q_T>30$ GeV$/c$, where our distribution is higher
than predicted by a factor of 1.25. However, the NNLO calculation agrees in shape with our data
when normalized at $q_T=30$ GeV$/c$.

%% file: acknowledgement_paragraph_r2.tex
% acknowledgement_paragraph_r2.tex                                 10/09/07
%
We thank C. Balazs, Q.H. Cao, P. Nadolsky, F. Petriello and C.P. Yuan
for many useful discussions.
We thank the staffs at Fermilab and collaborating institutions, 
and acknowledge support from the 
DOE and NSF (USA);
CEA and CNRS/IN2P3 (France);
FASI, Rosatom and RFBR (Russia);
CAPES, CNPq, FAPERJ, FAPESP and FUNDUNESP (Brazil);
DAE and DST (India);
Colciencias (Colombia);
CONACyT (Mexico);
KRF and KOSEF (Korea);
CONICET and UBACyT (Argentina);
FOM (The Netherlands);
Science and Technology Facilities Council (United Kingdom);
MSMT and GACR (Czech Republic);
CRC Program, CFI, NSERC and WestGrid Project (Canada);
BMBF and DFG (Germany);
SFI (Ireland);
The Swedish Research Council (Sweden);
CAS and CNSF (China);
Alexander von Humboldt Foundation;
and the Marie Curie Program.
%%%% remove Marie Curie at August 07 update
%